 \documentclass{ws-ijmpd}

\usepackage{graphicx}
\usepackage{epstopdf}
\usepackage{epsfig}
 \newcommand{\be}[1]{\begin{equation}\label{#1}}
 \newcommand{\ee}{\end{equation}}

\begin{document}
 \markboth{Shuang-Nan Zhang \& Yi Xie} {Neutron star magnetic field evolution produces timing noise}

%
\catchline{}{}{}{}{}
%

\title{Modeling pulsar time noise with long term power law decay modulated by short term oscillations of the magnetic fields of neutron stars}

\author{Shuang-Nan Zhang$^{1,2}$ \& Yi Xie$^1$}

\address{$^1$National Astronomical Observatories, Chinese Academy Of
Sciences, Beijing 100012, China.}
\address{$^2$Key Laboratory of Particle Astrophysics,
Institute of High Energy Physics, Chinese Academy of Sciences Beijing 100049, China.}

\maketitle

\begin{history}
\received{Day Month Year} \revised{Day Month Year} \comby{Managing Editor}
\end{history}

\begin{abstract}
We model the evolution of the magnetic fields of neutron stars as consisting of a long term power-law decay
modulated by short term small amplitude oscillations. Our model predictions on the timing noise $\ddot\nu$ of
neutron stars agree well with the observed statistical properties and correlations of normal radio pulsars.
Fitting the model predictions to the observed data, we found that their initial parameter implies their initial
surface magnetic dipole magnetic field strength $B_0\sim 5\times 10^{14}$~G when $t_0=0.4$~yr and that the
oscillations have amplitude $K\sim 10^{-8}$ to $10^{-5}$ and period $T$ on the order of years. For individual
pulsars our model can effectively reduce their timing residuals, thus offering the potential of more sensitive
detections of gravitational waves with pulsar timing arrays. Finally our model can also re-produce their
observed correlation and oscillations of $\ddot\nu$, as well as the ``slow glitch" phenomenon.
\end{abstract}

\keywords{Neutron Star; Magnetar; Magnetic Field Decay; Timing Noise.}

\section{Introduction}\label{sec1}

Many studies on the possible magnetic field decay of neutron stars have been done previously, based on the
observed statistics of their periods ($P$) and period derivatives ($\dot{P}$). Some of these studies relied on
population synthesis that includes complicated observational selection effects, unknown initial parameters of
neutron stars, and not completely understood models of radio emission of
pulsars\cite{Holt_1970,Bhattacharya_1992,Han_1997,Regimbau_2001,Gonthier_2002,Guseinov_2004,Aguilera_2008,Popov_2010};
consequently no firm conclusion can be drawn on if and how the magnetic fields of neutron stars decays (see
Refs.\cite{Harding_2006,Ridley_2010,Lorimer_2011}
for reviews). Alternatively some other studies used the spin-down or characteristics ages, $\tau_{\rm
c}=(P-P_{0})/2\dot{P}$ ($P_{0}$ is the initial spin period of the given neutron star), or $\tau_{\rm
c}=P/2\dot{P}$ if $P\gg P_{0}$, as indicators of the true ages of neutron stars and found evidence for their
dipole magnetic field decay\cite{Pacini_1969,Ostriker_1969,Gunn_1970}.
However, as we have shown recently, their spin-down ages are normally significantly larger than the ages of the
supernova remnants physically associated with them, which in principle should be the unbiased age indicators of
these neutron stars\cite{Lyne_1975,Geppert_1999,Ruderman_2005}.
It has been shown that this age mismatch can be understood if the magnetic fields of these neutron stars decay
significantly over their life times, since the magnetic field decay alters the spin-down rate of a neutron star
significantly, so that $T_{\rm SNR}/\tau_{\rm c}\ll 1$\cite{Lyne_1975,Geppert_1999,Ruderman_2005,zhang2011}.

Pulsars are generally very stable natural clocks with observed
steady pulses. However significant timing irregularies, i.e.,
unpredicted times of arrivals of pulses, exist for all pulsars
studied so far (see Ref.\cite{hobbs2010} for an extensive reviews of
many previous studies on timing irregularies of pulsars). The timing
irregularies of the first type are ``glitches", i.e., sudden
increases in spin rate followed by a period of relaxation; it has
been found that the timing irregularies of young pulsars with
$\tau_{\rm c}<10^5$ years are dominated by the recovery from
previous glitch events\cite{hobbs2010}. In many cases the neutron
star recovers to the spin rate prior to the glitch event and thus
the glitch event can be removed from the data satisfactorily without
causing significant residuals over model predictions. However in
some cases, glitches can cause permanent changes to both $P$ and
$\dot P$ of the neutron star, which cannot be removed from the data
satisfactorily. These changes can be modeled as permanent increase
of the surface dipole magnetic field of the neutron star;
consequently some of these neutron stars may grow their surface
magnetic field strength gradually this way and eventually have
surface magnetic field strength comparable to that of magnetars over
a life time of 10$^{5-6}$ years\cite{lin_highB}. These are the first
studies that linked the timing irregularies of pulsars with the long
term evolution of magnetic fields of neutron stars.

The timing irregularies of the second type have time scales of years
to decades and thus are normally referred to as timing noise. Hobbs
et al. carried out so far the most extensive study of the long term
timing irregularies of 366 pulsars\cite{hobbs2010}. Besides ruling
out some timing noise models in terms of observational
imperfections, random walks, and planetary companions, some of their
main conclusions are: (1) Timing noise is widespread in pulsars and
is inversely correlated with $\tau_{\rm c}$; (2) Significant
periodicities are seen in the timing noise of a few pulsars, but
quasi-periodic features are widely observed; (3) The structures seen
in the timing noise vary with data span, i.e., more quasi-period
features are seen for longer data span and the magnitude of
$|\ddot{\nu}|$ for shorter data span is much larger than that caused
by magnetic braking of the neutron star; and (4) The numbers of
negative and positive $\ddot{\nu}$ are almost equal in the sample.

Proper understanding of these important observations not only can lead to better understanding of the internal
structures of neutron stars, but can also be important for using some millisecond pulsars for gravitational wave
detections. In particular it is of high interests if some of the long term timing noise can be modeled and
removed reliably from data, thus improving the sensitivity of gravitational wave detections substantially. It is
extremely desirable if a physical model of neutron stars can catch and even predict the main features and long
term variations of the timing noise of pulsars, since such a model will ensure that no signal of gravitational
waves is removed when modeling the observed timing noise. Some of the observed periodicities have been suggested
as due to free precessions of neutron stars\cite{Stairs2000}. Some strong quasi-periodicities have been
identified recently as due to abrupt changes of the magnetospheric regulation of neutron stars, perhaps due to
varied particle emissions\cite{Lyne2010}. It was suggested that such varied magnetospheric particle emissions is
also responsible for the observed long term timing noise\cite{liu2010}. Alternatively it has been suggested that
the propagation of Tkachenko waves in a neutron star may modulate its rotational moment of inertia, and thus
produce the observed periodic or quasi-periodic timing noise\cite{Tkachenko66,Ruderman70,Haskell2011}. Because
none of the above processes may permanently change the properties of neutron stars, we can collectively
represent all the above modulations as some sort of oscillations of the observed surface magnetic field
strengths of neutron stars, as shown below.

In this work we first model the evolution of the dipole magnetic field of a neutron star as a long term
power-law decay modulated by one or several components of oscillations. We then calculate the spin-down history
of a neutron star within the magnetic braking model with the prescribed dipole magnetic field. The timing noise
is then calculated and compared with data. We show that all main observed features of neutron star's timing
noise can be reproduced satisfactorily. In the end we compare our model with other recently developed models of
neutron star's timing noise.

\section{Evolutionary model of the dipole magnetic field of neutron stars}\label{sec2}

Attributing the observed $T_{\rm SNR}/\tau_{\rm c}\ll 1$ to the long term decay of the magnetic field of a
neutron star\cite{Lyne_1975,Geppert_1999,Ruderman_2005,zhang2011}, we found that the magnetic field decay should
follow a power-law form, $B\propto t^{-\alpha}$ ($\alpha\approx0.5)$, in good agreement with the field decay
process dominated by the ambipolar diffusion mechanism\cite{Goldreich_1992,Heyl_1998}
with constant core temperatures\cite{zhang2011}. We further found that the core temperatures of magnetars,
normal radio pulsars and millisecond pulsars are approximately $10^8$ K, $10^7$ K, and $10^5$ K,
respectively\cite{zhang2011,xie2011}. The neutron star free precession, inferred from the observed periodic
timing noise\cite{Stairs2000}, is phenomenologically equivalent to periodic oscillation of its apparent surface
magnetic field $B=\sqrt{B_{\rm s}^{2}\sin^2\chi}$, where $\chi$ is the angle between the magnetic axis and spin
axis of the neutron and varies slightly as the neutron star precesses periodically. Similarly the observed
strong quasi-periodicities of timing noise can also be phenomenologically modeled as modulations of the observed
surface magnetic fields of these neutron stars, which are obtained from the observed $P$ and $\dot{P}$ of
neutron stars. We therefore model the evolution of the dipole magnetic field of a neutron star as a long term
power-law decay modulated by one or several components of oscillations,
\begin{equation}\label{b_decay}
B= B_{0}(\frac{t}{t_0})^{-\alpha}(1+\sum K_i\sin(\phi_i+2\pi\frac{t}{T_i})),
\end{equation}
where $t$ is the neutron star age, $t_0$ and $B_0$ are the starting time and initial surface dipole magnetic
field strength of the neutron star, $K_i$, $\phi_i$ and $T_i$ are the amplitude, phase and period of the
oscillating magnetic field of the $i$-th component, respectively.

Assuming pure magnetic dipole radiation as the braking mechanism for a pulsar's spin down, we have,
\begin{equation}\label{dipole}
I\Omega \dot \Omega  =-\frac{(BR^3)^2}{6c^3}{\Omega ^4},
\end{equation}
and
\begin{equation}\label{nu_dot}
\dot \nu  = -AB^2\nu^3,
\end{equation}
where $A=\frac{(2\pi R^3)^2}{6c^3I}$ is assumed to be a constant, $B$ is its dipole magnetic field at its
magnetic pole, $R$ is its radius, $I$ is its moment of inertia. We then have,
\begin{equation}\label{nu_ddot}
\ddot \nu  = -3AB^2\nu^2\dot \nu - 2AB\dot B\nu ^3.
\end{equation}
Simple calculations show that the value of the first term in the right part of the above equation is much less
than the observed $\ddot \nu$ as reported in Ref.\cite{hobbs2010}. Therefore the observed timing noise
characterized by $\ddot \nu$ may be dominated by the second term containing $\dot B$. From Eq.~(\ref{b_decay}),
we have,
\begin{equation}\label{bdot1}
\dot B= B(t)(-\frac{\alpha}{t}+\sum \frac{2\pi
K_i}{T_i}\cos(\phi_i+2\pi\frac{t}{T_i})).
\end{equation}
Assuming $K_i\ll1$ and ignoring the first term in Eq.~(\ref{nu_ddot}), we have,
\begin{equation}\label{nu_ddot2}
\ddot \nu\simeq -2\dot\nu(\frac{\alpha}{t}-\sum \frac{2\pi K_i}{T_i}\cos(\phi_i+2\pi\frac{t}{T_i})),
\end{equation}
where the first and second terms in the right part of the above equation are contributed by the long term
magnetic field decay and short term magnetic field oscillations, respectively. The age of the pulsar (not to be
confused with its characteristic or spin-down age $\tau_{\rm c}$) is given,
\begin{equation}\label{age}
t\simeq t_0(\frac{B_0}{B(t)})^{\frac{1}{\alpha}}.
\end{equation}
For a young pulsar with $t\lesssim 10^6$~yr, it is very likely that the first term in Eq.~(\ref{nu_ddot2})
dominates and thus it is generally anticipated that  $\ddot \nu>0$. For example, for $t\sim 5\times 10^3$~yr,
$\alpha \sim 0.5$, $P\sim 1$~s and $\dot \nu \sim 10^{-13}$~Hz~s$^{-1}$, the first term in Eq.~(\ref{nu_ddot2})
gives $\ddot \nu \sim 10^{-24}$~Hz~s$^{-1}$~s$^{-1}$, consistent with observations. On the other hand, the
second term in Eq.~(\ref{nu_ddot2}) dominates for an old pulsar, and thus both positive and negative $\ddot \nu$
should be observed with almost equal possibilities. This is in general agreement with
observations\cite{hobbs2010}.

\section{Statistical properties of the spin evolution of pulsars}

\begin{figure}

\begin{center}
{\includegraphics[width=10.0cm]{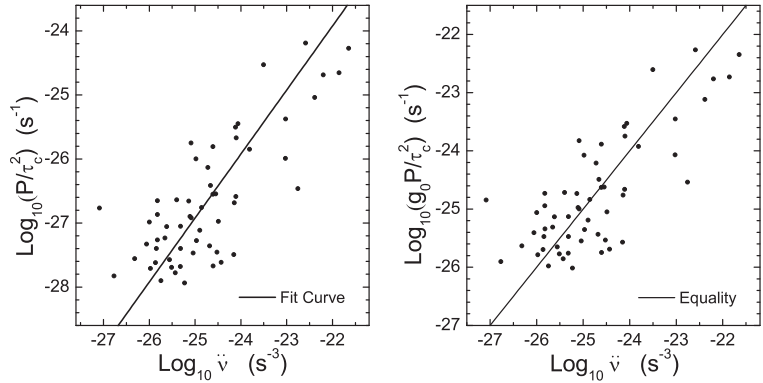} }
\end{center}
\caption{{\it Left}: Correlation of $\ddot \nu\sim P/\tau_{\rm c}^{2}$ for young pulsars with $\ddot \nu>0$ and
$\tau_{\rm c}<2\times 10^6$~yr. The solid line is the best fit with Eq.~(\ref{nu_ddot4}) ($f=0$), resulting in
$g_0=83.74$. {\it Right:} Correlation between the observed and model calculated $\ddot \nu$ for these young
pulsars, with $f=0$ in Eq.~(\ref{nu_ddot4}).}\label{f1}
\end{figure}

In estimating $\ddot \nu$ from observations,  $\nu$, $\dot \nu$, and $\ddot \nu$ are obtained by fitting the
phases of all pulses observed to the third order of its Taylor expansion over a period of time,
\begin{equation}\label{phase}
\Phi (t) = ({\Phi _0} + \nu t + \frac{1}{2}\dot \nu {t^2} + \frac{1}{6}\ddot \nu {t^3}).
\end{equation}
We can therefore estimate $\ddot \nu$ for a pulsar from
\begin{equation}\label{nu_ddot3}
\ddot \nu\sim -2\dot\nu(\frac{\alpha}{t}\pm f),
\end{equation}
where $f=2\pi\max({K_i}/{T_i})$. As shown later, $K_i\ll 1$ and $T_i$ is on the order of years, therefore $f\ll
1/t$ for young pulsars. Substitute Eq.~(\ref{age}) (with $\alpha=0.5$) into Eq.~(\ref{nu_ddot3}), our model of
power-law magnetic field decay in neutron stars predicts,
\begin{equation}\label{nu_ddot4}
\ddot \nu\sim g_0\frac{P}{\tau_{\rm c}^{2}}\pm \frac{f}{P \tau_{\rm c}},
\end{equation}
where $g_0={(3.2\times 10^{19})^2}/{(4t_0 B_{0}^{2})}$ is the initial parameter of pulsar's magnetic field. In
Fig.~\ref{f1} (left) we show the observed correlation of $\ddot \nu\sim P/\tau_{\rm c}^{2}$ for young pulsars
with $\ddot \nu>0$ and $\tau_{\rm c}<2\times 10^6$~yr. Fitting this correlation with that predicted by
Eq.~(\ref{nu_ddot4}) with $f=0$, we get $g_0=83.74$, which is considered to be a constant for all pulsars in the
rest of this paper, unless specified otherwise. Assuming $B_{0}=5\times10^{14}$~G, we have $t_0=0.39$~yr,
suggesting that these pulsars may be fast rotating magnetars when $t=t_0$. In Fig.~\ref{f1} (right) we show that
the observed and model calculated $\ddot \nu$ are very consistent with each other for these young pulsars, with
$f=0$ in Eq.~(\ref{nu_ddot4}); this is clear evidence for the power-law decay of neutron stars' magnetic fields,
at least for young pulsars.

\begin{figure}
\begin{center}
{\includegraphics[width=10.0cm]{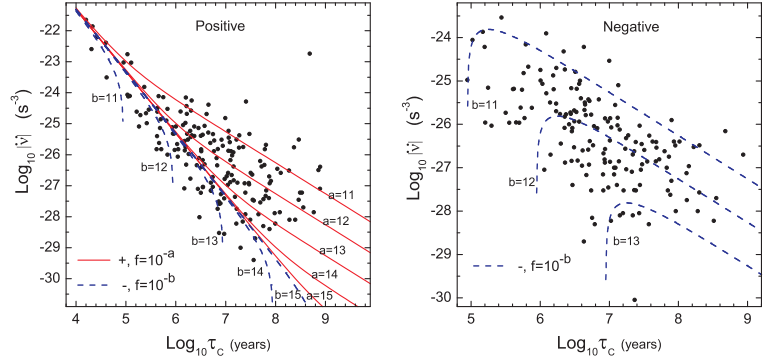} }
\end{center}
\caption{Correlation of $\ddot \nu\sim \tau_{\rm c}$ for all pulsars
with $\tau_{\rm c}< 10^9$~yr, i.e. recycled millisecond pulsars are
not considered. {\it Left}: pulsars with $\ddot \nu >0$. {\it
Right}: pulsars with $\ddot \nu <0$. The solid and dashed lines are
the model predicted correlation of $\ddot \nu\sim \tau_{\rm c}$ by
taking the `$+$' and `$-$' signs in Eq.~(\ref{nu_ddot4}),
respectively; different values of $f$ are labeled in the
figure.}\label{f2}
\end{figure}

\begin{figure}
\begin{center}
{\includegraphics[width=10.0cm]{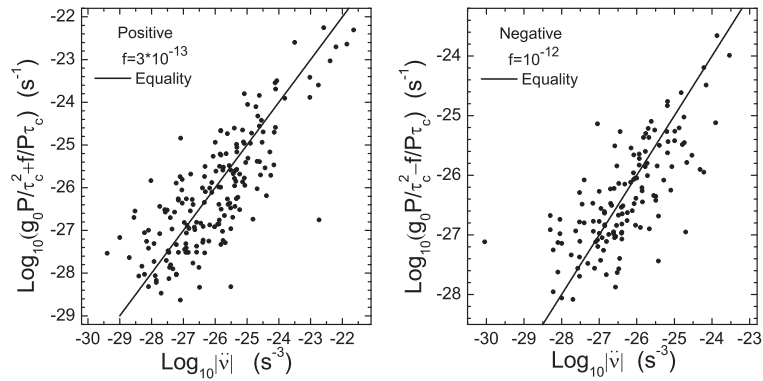} }
\end{center}
\caption{Comparison between the observed and model predicted $\ddot \nu$ with $f=3\times10^{13}$ or $10^{12}$
for $\ddot \nu>0$ or $<0$, respectively; here the observed value of $P$ of each pulsar is used in
Eq.~(\ref{nu_ddot4}).}\label{f3}
\end{figure}

In Fig.~\ref{f2} we show the observed correlation of $\ddot \nu\sim
\tau_{\rm c}$ for all pulsars with $\tau_{\rm c}< 10^9$~yr (i.e.
recycled millisecond pulsars are not considered); the left and right
panels displays all pulsars with $\ddot \nu >0$ and $\ddot \nu <0$,
respectively. Eq.~(\ref{nu_ddot4}) means that $\ddot \nu$ is
determined by both $P$ and $\tau_{\rm c}$, given $g_0$ and $f$.
Since the observed $P$ of these pulsars is narrowly distributed
around a median value of 0.6~s, therefore the observed main
correlation should be with $\tau_{\rm c}$, rather than $P$,
consistent with data\cite{hobbs2010}. In order to make
straightforward comparison between data and that predicted by
Eq.~(\ref{nu_ddot4}), we simply take $P=0.6$~s in
Eq.~(\ref{nu_ddot4}) to calculate the model predicted correlation of
$\ddot \nu\sim \tau_{\rm c}$ by choosing several values of $f$; the
calculations with the `$+$' and `$-$' signs in Eq.~(\ref{nu_ddot4})
are shown as solid and dashed lines, respectively. It should be
noted that the calculations with the `$-$' sign in
Eq.~(\ref{nu_ddot4}) can also produce positive $\ddot \nu$, and thus
dashed lines appear in both panels. For $f=10^{-13}$ and $T=3$~yr,
we have $K\approx10^{-6}$. In Fig.~\ref{f3} we show the comparison
between the observed and model predicted $\ddot \nu$. It is clear
that with just two parameters ($g_0$ is fixed and $f$ ranges between
$10^{-15}$ and $10^{-11}$), the general properties of the long term
spin evolution of all pulsars can be described successfully, except
for the recycled millisecond pulsars which we will address
separately. This can be considered as a strong support to our model
of the long term (power-law decay) and short term (small amplitude
oscillation) magnetic field evolution of neutron stars.

\section{Timing noise of individual pulsars}

Eq.~(\ref{nu_ddot4}) with fixed $g_0$ and $f$ cannot describe accurately the timing noise of individual pulsars
for at least the following reasons: (i) $g_0$ and $f$ may be different for different pulsars; (ii) multiple
oscillating components may exist with different phases, periods and amplitudes; (iii) the observed $\ddot \nu$
for each pulsar is determined by fitting the observed pulse phases with Eq.~(\ref{phase}) over a certain
observation time span $T_{\rm s}$; (iv) the observed $\ddot \nu$ correlates with $T_{\rm s}$, which cannot be
predicted by Eq.~(\ref{nu_ddot4}); (v) in using Eq.~(\ref{phase}) to fit data, all higher order terms are
absorbed in the four terms in Eq.~(\ref{phase}); and (vi) substantial residuals are often found in in using
Eq.~(\ref{phase}) to fit data. None of the last three issues can be addressed with Eq.~(\ref{nu_ddot4}). We
therefore should calculate numerically the history of a pulsar's pulse phase $\Phi(t)$ within our model and
compare it with observations.

The pulse phase change is given by
\begin{equation}\label{dphi}
d\Phi=(\nu+\dot\nu t)dt,
\end{equation}
where $\nu$ and $\dot \nu$ can be calculated numerically by combining Eqs.~(\ref{b_decay}) and (\ref{nu_dot}).
The entire history of $\Phi(t)$ for a pulsar can thus be calculated by integrating Eq.~(\ref{dphi}) numerically,
given the values of these parameters $t_0$, $B_0$, $\alpha$, $\phi_i$, $K_i$ and $T_i$. The calculated $\Phi(t)$
can then be fitted with Eq.~(\ref{phase}) to determine $\ddot \nu$ and residuals.

\subsection{Periodic and quasi-periodic residuals of timing noise}

\begin{figure}
\begin{center}
{\includegraphics[width=8.5cm]{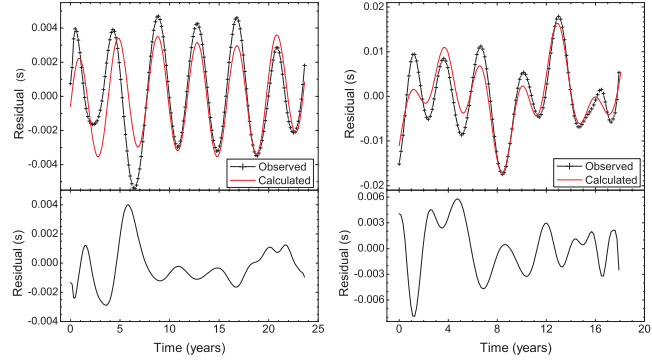} }
\end{center}
\caption{Residuals after fitting Eq.~(\ref{phase}) to the observed and calculated pulse phases for B1540$-$06
({\it left panel}) and B1826--17 ({\it right panel}), respectively. The dashed and solid lines in the top panels
are from the observed and calculated phases, respectively. The bottom panels are their differences.}\label{f6}
\end{figure}

For many pulsars the residuals in fitting data with
Eq.~(\ref{phase}) show period or quasi-periodic
behaviors\cite{hobbs2010}. From the FFT of the residuals of
B1540$-$06\cite{hobbs2010}, we can determine the phase and period of
the main peak in its power spectrum, which are taken as $\phi$ and
$T$ in Eq.~(\ref{b_decay}) for a single oscillation component. The
only remaining parameter is $K$. With different values of $K$, we
repeat the above procedure to calculate $\Phi(t)$, and fit it with
Eq.~(\ref{phase}) to determine residuals, which are then compared
with data until the minimum $\chi^2$ of the match is found.
Fig.~\ref{f6} (left) shows the comparison between the observed and
model predicted residuals with $K=7.41\times 10^{-10}$; the rms of
the residuals is reduced from $11.7$~ms to $7$~ms by introducing
just one parameter $K$, after subtracting the best fit residuals
from the observed ones.

When several quasi-periodic components are found in the residuals,
we take an iterative approach to find the phase and period of each
component: (1) find $\phi$, $T$ and the amplitude of the strongest
component in the FFT; (2) remove this periodic component from the
residuals; and (3) repeat the above until no significant component
is found in the FFT of the residual after all identified components
are subtracted. We then still use the $\chi^2$ optimization method
to find $K_i$ of each component. In Fig.~\ref{f6} (right) we show an
example for B1826--17, the FFT of the residuals of which has two
prominent components\cite{hobbs2010}; the rms of the residuals is
reduced from $33$~ms to $14$~ms in this case.

\subsection{Oscillating $\ddot \nu$ and slow glitches}

\begin{figure}
\begin{center}
{\includegraphics[width=8.5cm]{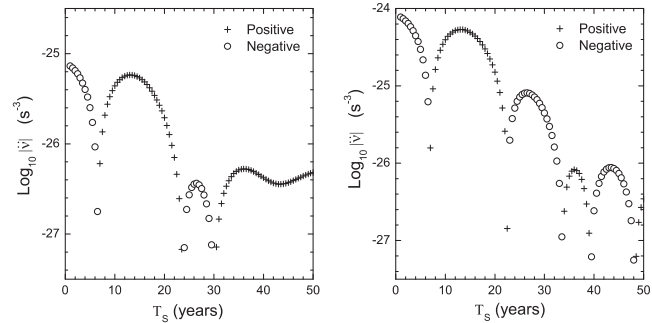} }
\end{center}
\caption{{\it Left}: Calculated correlation of $\ddot \nu\sim T_{\rm s}$ for B0329$+$54, with one oscillation
component of $T=15.5$~yr and $K=10^{-8}$. {\it Right}: The same as the left panel, except that $K=10^{-7}$.
}\label{f4}
\end{figure}

We then try to model the timing noise of B0329$+$54, which shows $\ddot \nu$ correlates with $T_{\rm s}$ and
even switches between positive and negative values; the parameters we choose are given in the caption of
Fig.~\ref{f4}, the left panel of which shows a similar correlation of $\ddot \nu\sim T_{\rm s}$ in Fig.~12 of
Ref.\cite{hobbs2010}. This shows that a single pulsar can produce $\ddot \nu$ over almost the entire range of
$\ddot \nu$ for all pulsars, depending on the chosen observation time span $T_{\rm s}$; this may be the main
source of the observed wide scatter in Fig.~\ref{f2}. In Fig.~\ref{f4} (right), the calculated correlation of
$\ddot \nu\sim T_{\rm s}$ is shown with $K=10^{-7}$, exhibiting clear oscillations of $\ddot \nu$ between
positive and negative values. This explains naturally the almost equal numbers of positive and negative values
of $\ddot \nu$ reported\cite{hobbs2010}.

\begin{figure}
\begin{center}
{\includegraphics[width=8.5cm]{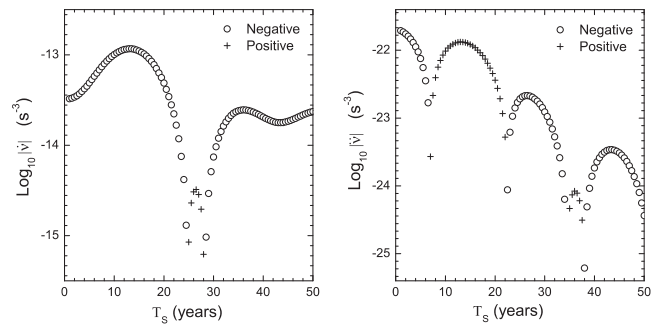} }
\end{center}
\caption{Calculated correlations of $\dot \nu\sim T_{\rm s}$ ({\it left panel}) and $\ddot \nu\sim T_{\rm s}$
({\it right panel}) with the following parameters: $T=15.5$~yr and $K=2.5\times 10^{-5}$.}\label{f5}
\end{figure}

We then increase $K$ to $K=2.5\times 10^{-5}$ and show in Fig.~\ref{f5} the calculated correlations of $\dot
\nu\sim T_{\rm s}$ and $\ddot \nu\sim T_{\rm s}$. We find that sometimes $\dot \nu$ switches to positive values,
quite similar to the recently found slow glitches\cite{zou2004,Shabanova2007}; the observed change of sign of
$\ddot \nu$ is also consistent with our model prediction. However the calculated $\ddot \nu$ becomes negative
most of the times. This means that the phenomenon of slow glitches may just be a manifestation of unusually
large amplitude oscillations of the apparent magnetic field of a neutron star, consistent with the observed
unusually large rms of residuals for the two pulsars observed with slow glitches, i.e.,  1284.3~ms and 613.7~ms
for PSR~J1825$-$0935 (B1822$-$09) and PSR~J1835$-$1106, respectively.
\section{Conclusions and discussions}
Our previous work has found evidence that the magnetic fields of neutron stars have long term decays in a
power-law form, agreeing with the prediction of the ambipolar diffusion mechanism. Through examining the
observed (apparent) short term variations of the magnetic fields of neutron stars, we propose that the
(apparent) magnetic fields of neutron stars have oscillations. We thus model the evolution of a neutron star's
magnetic field as consisting of a long term power-law decay modulated by short term small amplitude
oscillations. Our model predicted $\ddot\nu$, a parameter widely used to characterize the timing noise of
pulsars, agrees well with the observed statistical properties and correlations of all pulsars (except those
recycled millisecond pulsars). This can be considered as a strong support to our model of magnetic field
evolution of neutron stars.

Fitting the model predictions to the observed data, we found that their initial parameter implies their initial
surface dipole magnetic field strength $B_0\sim 5\times 10^{14}$~G when $t_0=0.4$~yr and that the oscillations
have amplitude $K\sim 10^{-8}$ to $10^{-5}$ and period $T$ on the order of years. We then calculated the timing
properties of individual pulsars with our model. We modeled the observed timing residuals of two pulsars and
demonstrated that our model can effectively reduce their timing residuals, thus offering the potential of more
sensitive detections of gravitational waves with pulsars. Depending on the combination $K$ and $T$, we can
re-produce the observed correlation and oscillations of $\ddot\nu$, as well as the ``slow glitch" phenomenon.

We did not study the timing noise properties of the recycled millisecond pulsars with our model, because their
initial properties may be substantially different from the normal radio pulsars we studied in this work.

\section*{Acknowledgments}
 We thank interesting discussions with Renxin Xu on timing noise. SNZ acknowledges partial
funding support by the National Natural Science Foundation of China under project no. 11133002, 10821061,
10725313, and by 973 Program of China under grant 2009CB824800.

\end{document}